\documentclass[aps,prl,twocolumn,amsmath,amssymb,superscriptaddress]{revtex4}
\usepackage{graphicx}
\usepackage{amsmath}
\usepackage{hyperref}

\begin{document}

\title{Paired Superfluidity and Fractionalized Vortices in Spin-orbit Coupled Bosons}
\author{Chao-Ming Jian}
\email{cm.jian.physics@gmail.com}
\affiliation{Institute for Advanced Study, Tsinghua University, Beijing, 100084, China}
\author{Hui Zhai}
\email{hzhai@mail.tsinghua.edu.cn}
\affiliation{Institute for Advanced Study, Tsinghua University, Beijing, 100084, China}
\date{\today}
\begin{abstract}
In this letter we study finite temperature properties of spin-$1/2$ interacting bosons with spin-orbit coupling in two dimensions. When the ground state has stripe order, we show that thermal fluctuations will first melt the stripe order and lead to a superfluid of boson pairs if the spin-orbit coupling is isotropic or nearly isotropic. Such a phase supports fractionalized quantum vortices. The Kosterlize-Thouless transition from superfluid to normal state is driven by proliferation of half vortices. When the ground state is a plane wave state, the transition to normal state is driven by conventional Kosterlize-Thouless transition. However, the critical temperature will drop to zero for isotropic spin-orbit coupling.
\end{abstract}

\maketitle

Spin-orbit coupling (SOC) plays an important role in a wide range of quantum systems from condensed matter, atomic physics to nuclear physics. For example, recent studies of electronic systems with strong SOC have led to the discovery of  topological insulators \cite{TBI} with potential applications to new quantum devices. Recently, bosons with SOC have also begun to attract considerable attentions \cite{SObec,SObecHo,BEC,Yip,spin2}, for such a coupling can now be generated in cold atom systems by engineering atom-light interaction \cite{Dalibard}. Very recently, a pioneer experiment in NIST has already achieved boson condensate with a restricted class of SOC  \cite{IanNonAbelian}. Mean-field studies of the ground state have revealed two different phases, i.e. the plane wave phase and the stripe phase \cite{SObec,SObecHo,Yip,IanNonAbelian}. However, the fluctuation effects on top of mean-field description and the finite temperature properties have not been studied yet. In this letter we show that finite temperature thermal fluctuations will indeed lead to new phase and phase transitions.

In this work we consider spin-$1/2$ interacting bosons with SOC in two dimensions, and present a global phase diagram of this system in terms of the interaction parameters, temperature and the degree of anisotropy of SOC. The main results are summarized as follows:

{\bf (1)} For stripe phase with isotropic or nearly isotropic SOC, the system undergoes two phase transitions as temperatures increases. Stripe order melts at the lower transition temperature and superfluidity disappears at the higher one. Between these two transitions, bosons form pairs and these boson pairs exhibit superfluidity. In this unusual boson paired phase, vortices are fractionalized.

{\bf (2)} For the stripe phase with anisotropic SOC, the system undergoes a direct Kosterlize-Thouless (KT) transition to normal state. However, since the energy of a half vortex bound to a single dislocation is lower than that of a usual vortex, the KT transition is always driven by proliferation of half vortices.

{\bf (3)} For the plane wave phase, the system undergoes a conventional KT transition to normal phase driven by proliferation of usual vortices. However, the KT temperature drops to zero for isotropic SOC.

We note such a boson paired phase and unconventional KT transition have also been proposed in other systems with different microscopic physics, such as stripe phase in high-Tc superconductor materials \cite{Kivelson}, ``Fulde-Ferrell-Larkin-Ovchinnikov" state \cite{PDW} of superconductor in Zeeman field and spin-1 bosons \cite{Fei}, however, it has not been realized experimentally yet. Ultracold bosons with SOC provide a new route toward this interesting physics, and may shed lights on understanding other systems.

The single particle Hamiltonian $\mathcal{H}_0$ with SOC is
\begin{equation}
\mathcal{\hat{H}}_0=\int d^2{\bf r}\left\{
\hat{\psi}^{\dag}\left[\frac{\hat{\bf p}^2}{2m}-\frac{1}{m}(\kappa_x\sigma_x\hat{p}_x+\kappa_y\sigma_y\hat{p}_y)\right]\hat{\psi}
\right\}. \label{GLfreeEnergy2}
\end{equation}
where $\hat{\psi}=\left(\begin{array}{c}\hat{\psi}_0 \\ \hat{\psi}_1\end{array}\right)$ is a two-component spinor, and $\sigma_x,\sigma_y$ are Pauli matrices. Without loss of generality, we can define $\kappa_x=\kappa$ and $\kappa_y=\eta\kappa$ with $0\leqslant \eta \leqslant 1$. In this case, the single particle energy minimum is located at ${\bf k}=\pm \kappa \hat{x}$ with their wave function given by $\varphi_{\pm }=e^{\pm i\kappa x}\left(\begin{array}{c}1 \\  \pm 1\end{array}\right)$.
For interactions, generally speaking it can be written as
\begin{equation}
\mathcal{\hat{H}}_{\text{int}}=\int d^2{\bf r}\left(g_{00}\hat{n}^2_{0}+g_{11}\hat{n}^2_{1}+2g_{01}\hat{n}_{0}\hat{n}_{1}\right).
\end{equation}
In Ref. \cite{SObec} we considered the mean-field ground state of a simplified situation with $g_{00}=g_{11}=g_0$, where $\mathcal{\hat{H}}_{\text{int}}$ can be rewritten as
\begin{equation}
\mathcal{\hat{H}}_{\text{int}}=\frac{1}{2}\int d^2{\bf r} \left((2g_0+g)\hat{N}^2-g \hat{S}^2_{z}\right),
\end{equation}
where $\hat{N}=\hat{n}_0+\hat{n}_1$, $\hat{S}_z=\hat{n}_0-\hat{n}_1$, and $g=g_{01}-g_{00}$.
We found two distinct phases for $g>0$ or $<0$. For $g<0$, all bosons condense into a spontaneously chosen single plane wave state, either $\varphi_{+}$ or $\varphi_{-}$. This state breaks time-reversal symmetry. It is named  ``plane wave phase". For $g>0$, bosons condense into a superposition state of these two plane wave states. Consequently, this state exhibits stripe order of spin density, and is named ``stripe phase". Later studies show that this conclusion will not change qualitatively even if one considers more complicated interactions \cite{Yip}. These two phases are analogous to ``Fulde-Ferrell" or ``Larkin-Ovchinnikov" states proposed for superconductor in a Zeeman field.

\begin{figure}[tbp]
\includegraphics[height=2.in, width=2.in]
{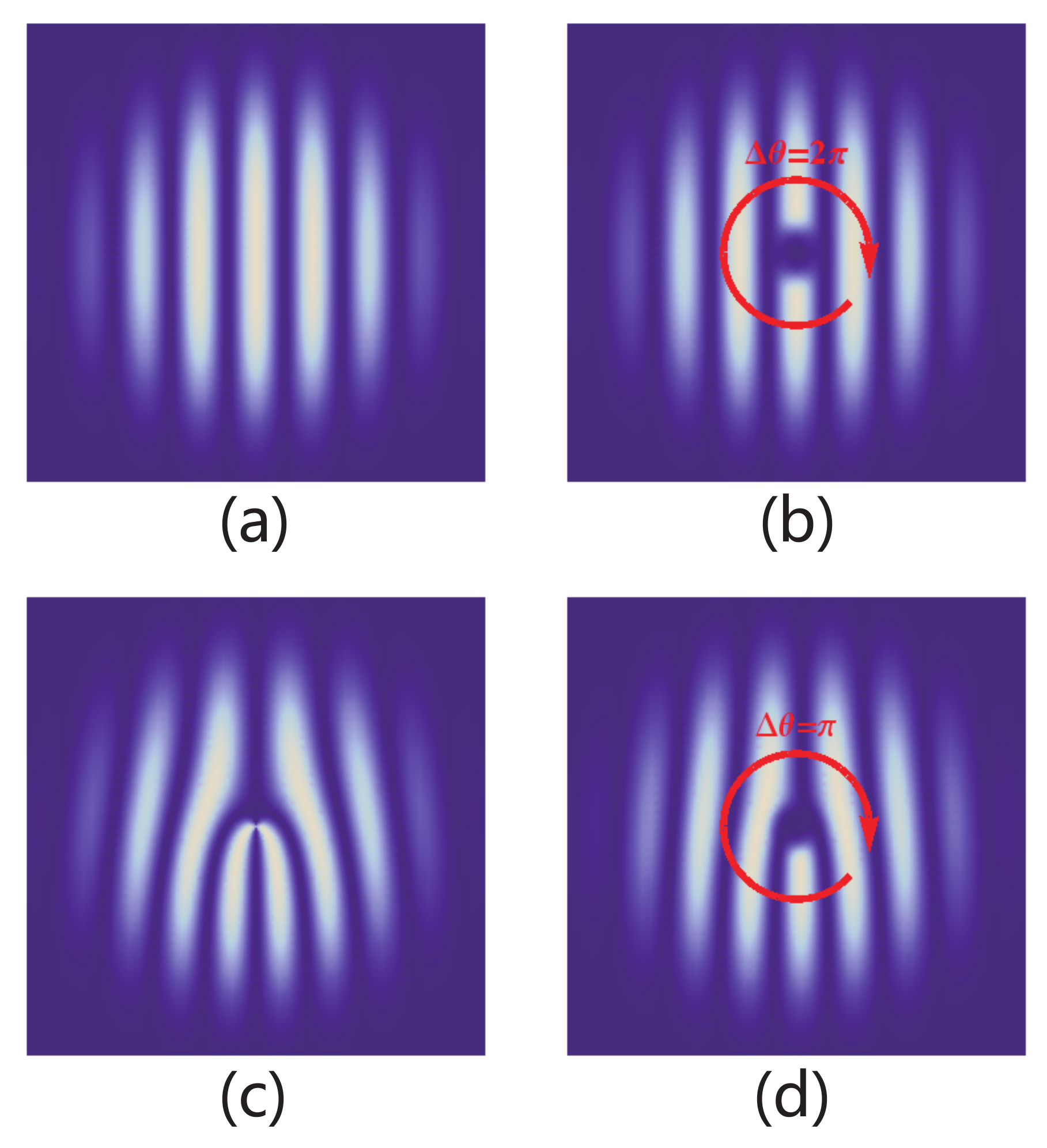}\caption{The amplitude of the order parameter with (a) no topological defect, (b) a single vortex, (c) double dislocations and (d) a half vortex bound to a dislocation. The red arrow indicates that the superfluid phase changes $2\pi$ around a single vortex (b) and $\pi$ around a half vortex bound with a dislocation (d). \label{TopoDef}}
\end{figure}

For a typical situation in cold atoms,  $g_0\gg |g|$. Thus, we take the approximation that only spin and phase fluctuations are considered in the low-energy theory and the density fluctuations are ignored. First, let us consider the stripe phase, and we introduce the ansatz as
\begin{align}
\varphi=\frac{\sqrt{\rho_s}}{2}\left[e^{i \kappa x+i\theta_{+} }\left(\begin{array}{c}e^{\frac{-i\xi_{+}}{2}}\\e^{\frac{i\xi_{+}}{2}}\end{array}\right)  +e^{-i\kappa x+i\theta_{-}}\left(\begin{array}{c}e^{\frac{-i\xi_{-}}{2}}\\ -e^{\frac{i\xi_{-}}{2}}\end{array}\right)
\right]. \nonumber
\end{align}
$\theta\equiv(\theta_{+}+\theta_{-})/2$ and $u\equiv(\theta_{+}-\theta_{-})/2$ describe two low-lying degrees of freedom. $\theta$ is the total phase which corresponds to the superfluid phonon, and $u$ is the relative phase between two counter propagating waves, which corresponds to the phonon mode of the stripe. The requirement of single-valuedness of the wave function $\varphi$ imposes a constraint that the winding numbers of $\theta_{+}$ and $\theta_{-}$ must be integers. In terms of $\theta$ and $u$, that means $q_{\text{v}}\pm q_{\text{d}}$ are both integers, where $q_{\text{v}}$ and $q_{\text{d}}$  denote the topological charge of vortices (defects in $\theta$) and dislocations (defects in $u$), respectively. Three distinct elementary topological defects can be identified: (i) a full vortex $(q_{\text{v}},q_{\text{d}})=(\pm1,0)$ (Fig. \ref{TopoDef}(b)) (ii) double dislocations $(q_{\text{v}},q_{\text{d}})=(0,\pm1)$ (Fig. \ref{TopoDef}(c)) and (iii) a half vortex bound to a single dislocation $(q_{\text{v}},q_{\text{d}})=(\pm1/2,\pm1/2)$ (Fig. \ref{TopoDef}(d)). In two dimensions, proliferation of these defects will play crucial roles in finite temperature KT transitions.

For the plane wave phase, superfluid phase is the only low-energy degree of freedom, and one can choose the ansatz as
\begin{equation}
\varphi=\sqrt{\frac{\rho_s}{2}}e^{i\kappa x+i\theta}
\left(\begin{array}{c}
e^{-i\xi/2}\\
e^{i\xi/2}
\end{array}\right).\label{PWansatz}
\end{equation}
In this case superfluid vortex is the only type of topological excitation.

Using these ansatz, one can obtain the effective low-energy Hamiltonian following three steps: (i) First one can obtain an energy functional $\mathcal{F}=\langle \varphi|\hat{H}_0-g\int d^2{\bf r}\hat{S}^2_{z}/2|\varphi\rangle$, which characterizes the fluctuation of $\theta$, $u$, $\xi_{+}$ and $\xi_{-}$ for the stripe phase, or the fluctuation of $\theta$ and $\xi$ for the plane wave phase. (ii) By expanding it to quadratic orders of these fields, we note that both $\xi_{+}$ and $\xi_{-}$ (or $\xi$) are massive modes; and (iii) one can then integrate out $\xi_{+}$ and $\xi_{-}$ (or $\xi$), which yields an effective Hamiltonian for the low-energy physics. The calculations are lengthy but quite straightforward, and we present all the details in the supplementary material \cite{Supplementray}. Hereafter, we shall discuss the results for different cases:

{\it Stripe Phase with Isotropic SOC:} We first consider isotropic SOC with $\eta=1$. In this case, the Hamiltonian has rotational symmetry and the single particle energy minimum is in fact
a circle in momentum space with $|{\bf k}|=\kappa$. Nevertheless, the single plane wave state and the stripe state are still the only two phases, because the repulsive density-density interaction $(2g_0+g)\hat{N}^2$ term does not favor density modulation, and therefore rule out a superposition state of more than two plane wave states. (Unless some other interaction terms are included into the Hamiltonian such as for spin-$2$ bosons \cite{spin2}, which is beyond the scope of this work.)

In this case, the stripe phase can also been called ``smectic superfluid" in terms of the language of liquid crystal physics, because it breaks rotational symmetry and translation symmetry along one spatial direction as shown in Fig. \ref{TopoDef}(a). Following quite straightforward derivations \cite{Supplementray} we will arrive at a low-energy effective Hamiltonian as
\begin{align}
\mathcal{H}_{\text {eff}}^{{\text ST}}=\frac{\rho_s}{2m}\left[\left(\partial_x\theta\right)^2+\frac{\left(\partial_y\theta\right)^2}{\alpha^2}+
 \left(\partial_x u\right)^2+\frac{\left(\partial_y^2 u \right)^2}{4\kappa^2}
\right], \label{SWHeff}
\end{align}
where $\alpha=\sqrt{1+2\kappa^2/(m\rho_s g)}$.
The most interesting feature in $\mathcal{H}_{\text {eff}}^{{\text ST}}$ is the absence of $(\partial_y u)^2$ term, which is a manifestation of rotational symmetry. Similar effective energy has also been found in the classical smectic liquid crystal \cite{Lubensky}. A simple argument can be given as follows: consider a small fluctuation of $u=qy$, i.e. $\theta_{+}=qy$ and $\theta_{-}=-qy$, from the ansatz $\varphi$, it is clear that this in fact corresponds to a change of stripe direction from $\hat{x}$ to $\kappa \hat{x}+q\hat{y}$, and the module of wave vector is also changed by $\sqrt{q^2+\kappa^2}-\kappa \propto q^2$, hence, the increase of kinetic energy is proportional to $q^4$ instead of $q^2$.

In the derivation of the effective theory, it is important to keep two massive modes $\xi_{+}$ and $\xi_{-}$ first, which represent the spin degrees of freedom in the left and right moving components, and then integrate them out. Otherwise, one will end up with a wrong effective Hamiltonian as two decoupled isotropic $XY$ models. This is because the rotational symmetry in this system is in fact a simultaneous rotation of both spin and space.

Since the superfluid phase is governed by an anisotropic {\it XY} model, an algebraic order is expected at low temperature. However, the stripe order behaves in a very different way due to the absence of the second order derivative term along $\hat{y}$. The correlation of stripe order is given by
$\langle e^{iu({\bf r})}e^{-
iu(0)}\rangle=e^{-\frac{1}{2}\langle [u({\bf r})-u(0)]^2\rangle}$
and
\begin{eqnarray}
\langle [u({\bf r})-u(0)]^2\rangle=\frac{2m}{\beta\rho_s}\int \frac{d^2{\bf q}}{(2\pi)^2}\frac{1-e^{i{\bf q}{\bf r}}}{q_x^2+q_y^4/(4\kappa^2)}.
\end{eqnarray}
Along $\hat{x}$-direction, the integral becomes
\begin{align}
\frac{\sqrt{|x|}}{\pi^2}\int^\infty_0 d \tilde{q}_x \int^{\infty}_{0} d \tilde{q}_y\frac{1-\cos{\tilde{q}_x}}{\tilde{q}^{3/2}_x}
\frac{1}{1+\tilde{q}_y^4/(4\kappa^2)}=\sqrt{\frac{\kappa|x|}{2\pi}}\nonumber
\end{align}
where $\tilde{q}_x=q_x|x|$ and $\tilde{q}_y=q_y/\sqrt{|q_x|}$. And along $\hat{y}$-direction, the integral becomes
\begin{align}
|y|\int \frac{d \tilde{q}_x d \tilde{q}_y}{(2\pi)^2}\frac{1-e^{i\tilde{q}_y}}{\tilde{q}^{2}_y}
\frac{1}{1/(4\kappa^2)+\tilde{q}_x^2}=\frac{\kappa|y|}{2}. \nonumber
\end{align}
where $\tilde{q}_x=q_x/q^2_y$ and $\tilde{q}_y=q_y|y|$.
Hence, the correlation of stripe order decays exponentially as
\begin{equation}
\langle e^{iu({\bf r})}e^{-
iu(0)}\rangle\propto{\begin{cases}e^{-m/(\beta\rho_{s})\sqrt{\kappa|x|/(2\pi)}} & \text{along $\hat{x}$ axes } \\ e^{-m\kappa |y|/(2\beta\rho_s)}& \text{along $\hat{y}$ axes } \end{cases}}, \label{DWcorrel}
\end{equation}

\begin{figure}[tbp]
\includegraphics[height=1.5in, width=2.5in]
{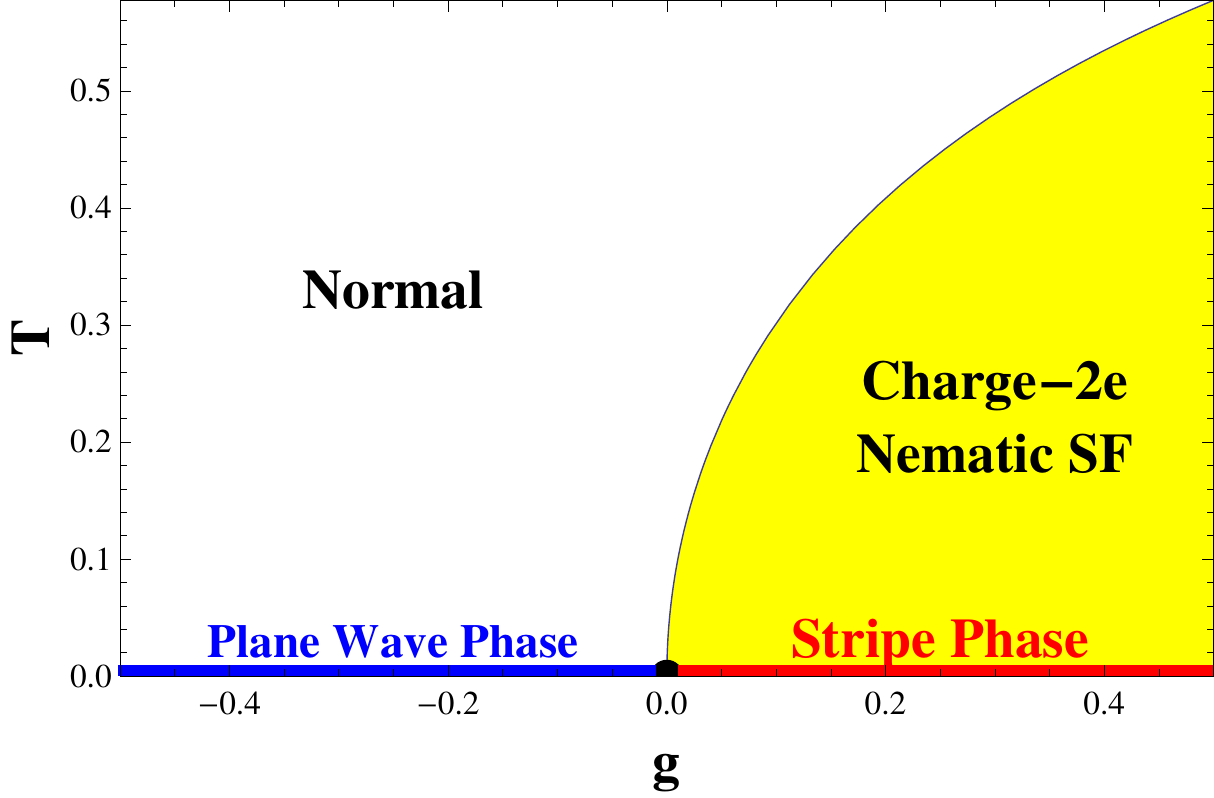}\caption{The $g-T$ phase diagram of two-dimensional bosons with isotropic SO coupling. $g$ is in the unit of $2\kappa^2/(m\rho_s)$ , and $T$ is in the unit of $\pi\rho_s/(2m)$ \label{RBPD}}
\end{figure}

Eq. (\ref{DWcorrel}) shows that the stripe order becomes disordered at any finite temperature. Hence, the broken translation symmetry along the stripe direction is restored, and the ``smectic superfluid" turns into a ``nematic superfluid", where only rotational symmetry is broken. Moreover, because the relative phase between two components is disordered, single boson field can not have broken symmetry, i.e. $\langle \hat{\psi}\rangle=0$. However, boson pair field $\langle\hat\psi\hat{\psi}\rangle$ does not couple to the relative phase, and therefore, still exhibits (quasi)-long range order until their superfluid phase $2\theta$ undergoes KT transition. This phase is named ``charge-2e nematic superfluid" in the phase diagram Fig. \ref{RBPD}, where ``charge" means particle number.

This behavior can also been understood from the aspect of topological defects. Because of the absence of $(\partial_y u)^2$ term, the energy of double dislocations $(0,\pm 1)$ does not have logarithmic divergence, and fails to compete with entropy. Therefore dislocations proliferate at any finite temperature and melt the stripe order. And also, with $u$ field disordered, a $(\pm 1/2,\pm 1/2)$ defect essentially becomes a half vortex, and its energy is much smaller than a single $(\pm 1,0)$ vortex. Hence, the superfluid transition of paired bosons is driven by proliferation of half vortices. By comparing its energy to entropy, one can roughly estimate the transition temperature as $T^c=\pi\rho_s/(8m\alpha)$, which is reduced by a factor of $1/(4\alpha)$ compared to conventional KT transition.

\begin{figure}[tbp]
\includegraphics[height=1.5in, width=2.5in]
{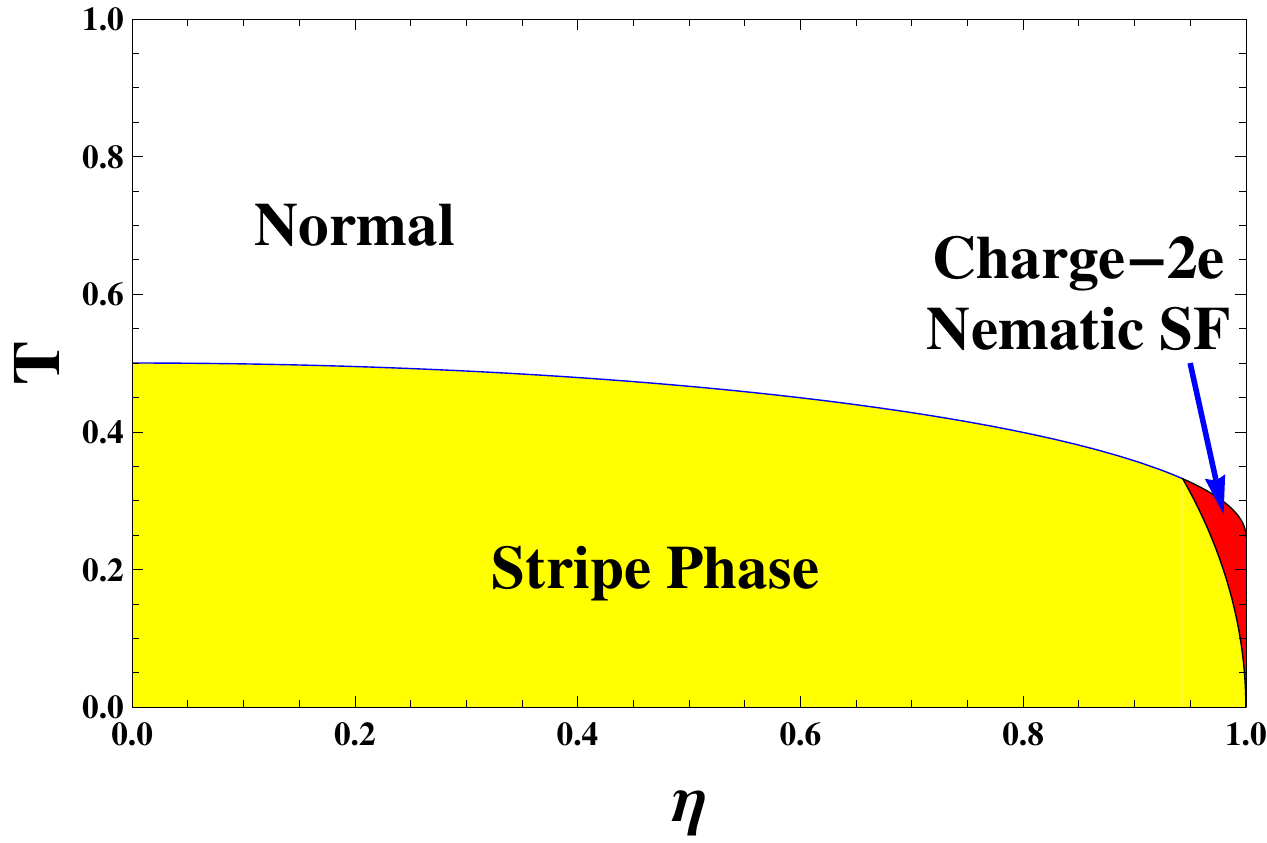}\caption{The $\eta-T$ phase diagram of two-dimensional bosons with $g>0$, where $\eta=\kappa_y/\kappa_x$ represents the degree of anisotropy of SO coupling. $T$ is in the unit of $\pi\rho_s/(2m)$ \label{Phase_diagram_eta}}
\end{figure}

{\it Plane Wave Phase with Isotropic SOC:} Following similar derivations \cite{Supplementray}, we arrive at an effective Hamiltonian for the plane wave phase as
\begin{equation}
\mathcal{H}_{\text {eff}}^{^{\text PW}}=\frac{\rho_s}{2m}\left[\left(\partial_x\theta\right)^2+\frac{1}{4\kappa^2}(\partial_y^2\theta)^2\right]
.\label{PWHeff}
\end{equation}
Here the consequence of rotational symmetry breaking is now manifested as the absence of $(\partial_y\theta)^2$ term, which can be understood using similar argument presented above. Similar analysis can show that the superfluid phase will disorder at any finite temperature. Hence, we have completed the $g-T$ phase diagram for isotropic SO coupling as shown in Fig. \ref{RBPD}.

{\it Stripe Phase with Anisotropic SOC:} When $\eta<1$, the rotational symmetry is broken explicitly. As expected, $(\partial_y u)^2$ no longer vanishes in the low-energy effective Hamiltonian given below
\begin{align}
\mathcal{H}^{\text {ST}}_{\text{eff}}=\frac{\rho_s}{2m}\left\{(\partial_x\theta)^2+\left[(1-\eta^2)+\frac{\eta^2}{\alpha^2}\right](\partial_y\theta)^2 \nonumber
\right.\\
\left.+(\partial_x u)^2+(1-\eta^2)(\partial_y u )^2\right\}, \label{effST}
\end{align}
and the higher derivative terms can now be ignored. However, the resulting ${\it XY}$ model is very anisotropic. In particular, the coefficient of $(\partial_y u)^2$ term has to vanish when $\eta\rightarrow 1$. For a two-dimension ${\it XY}$ model, phase transitions are driven by proliferation of different types of topological excitations. By calculating the energy of different type of topological excitations from the effective Hamiltonian Eq. \ref{effST}, we can obtain the characteristic temperatures for proliferation of different topological defects
\begin{equation}
T_{(q_v,q_d)}=\frac{\pi\rho_s}{2m}\left[\sqrt{(1-\eta^2)+\frac{\eta^2}{\alpha^2}}q_v^2+\sqrt{1-\eta^2}q_d^2\right].\label{AniTransT}
\end{equation}
It is easy to see that $T_{(\pm 1, 0)}$ is always higher than $T_{(\pm 1/2,\pm 1/2)}$. And for $\eta>\eta_{\text c}=\sqrt{8\alpha^2/(8\alpha^2+1)}$, $T_{(\pm 1/2,\pm 1/2)}>T_{(0,\pm 1)}$. At $T_{(0, \pm 1)}$, proliferation of dislocations first melts the stripe order, but does not destroy superfluid phase. Therefore, the system enters ``charge-2e nematic superfluid" phase. Then, at $T_{(\pm 1/2,\pm 1/2)}$, half vortices proliferate and the system becomes normal. For $\eta<\eta_{\text c}$,  $T_{(\pm 1/2,\pm 1/2)}<T_{(0,\pm 1)}$, $(\pm 1/2,\pm 1/2)$ defects proliferate first at $T_{(\pm 1/2,\pm 1/2)}$, destroying both superfluid and stripe order simultaneously, and driving the system into normal phase directly. The $\eta-T$ phase diagram in the regime $g>0$ is shown in Fig. \ref{Phase_diagram_eta}.

\begin{figure}[tbp]
\includegraphics[height=1.7 in, width=1.8in]
{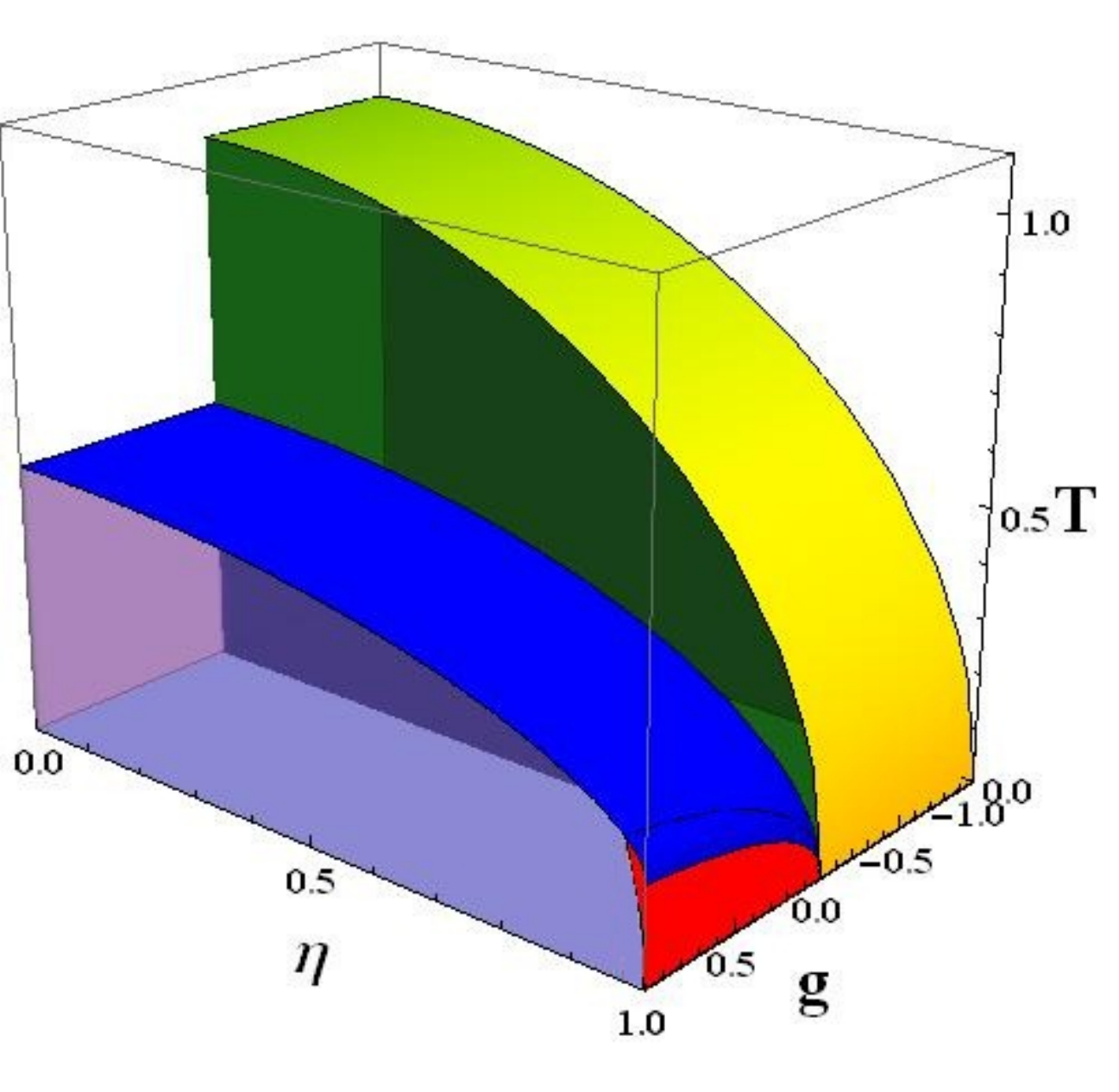}\caption{The $g-\eta-T$ phase diagram of two-dimensional bosons with SO coupling. Below the blue surface, on the left side of red surface, it is stripe ordered superfluid, and on the right side of red surface, it is charge-2e nematic superfluid; Below the yellow surface it is plane wave phase. $g$ is in the unit of $2\kappa^2/(m\rho_s)$ , and $T$ is in the unit of $\pi\rho_s/(2m)$ \label{Phases}}
\end{figure}

{\it Plane Wave Phase with Anisotropic SOC:} For the plane wave phase at $g<0$, the effective Hamiltonian can be derived as
\begin{equation}
\mathcal{H}^{\text{PW}}_{\text{eff}}=\frac{\rho_s}{2m}\left[(\partial_x\theta)^2+(1-\eta^2)(\partial_y\theta)^2\right]
\label{HeffPw}
\end{equation}
Similarly, the coefficient of $(\partial_y\theta)^2$ vanishes only when $\eta\rightarrow 1$. In this regime superfluid will turn into normal at $T=(\pi\rho_s)\sqrt{1-\eta^2}/(2m)$, and the phase transition is driven by integer vortices which is the only type of topological excitations in this regime. The transition temperature is higher than the KT transition driven by half vortices in the stripe phase.

Hence, we have reached a global phase diagram for this system as shown in Fig. \ref{Phases}. Our results can be generalized to three dimensions. One should also find that stripe order melts before superfluid phase is disordered, which allows a paired superfluid in three dimensions. Finally, we note that, in the real situation, there may be small difference between $g_{00}$ and $g_{11}$, and in some cases there is also a Zeeman field $h\sigma_z$ term. They will will not affect the nature of these two phases, and also will not affect the fluctuation analysis discussed above.

Main predictions of this work can be verified experimentally. For instance, recently various techniques have been applied to study KT transition for spinless bosons; Melting of stripe order can be detected through Bragg spectroscopy or in-situ image of density. Different topological defects can also been differentiated through interference technique. Boson pairing may manifest itself in the high-order correlation functions. Our results provide insights for future experimental study in this new system.

{\it Acknowledgment:} We thank Chushun Tian, Jung Hoon Han and Lee Chang for carefully reading our manuscript. This work is supported by Tsinghua University Initiative Scientific Research Program, NSFC under Grant No. 11004118 and NKBRSFC under Grant No. 2011CB921500.

\begin{widetext}

{\it Appendix:} In this supplementary material, we present some details of derviation of the low-energy effective Hamiltonian.

For the stripe phase, we have introduced the ansatz
\begin{align}
\varphi=\frac{\sqrt{\rho_s}}{2}\left[e^{i \kappa x+i\theta_{+} }\left(\begin{array}{c}e^{\frac{-i\xi_{+}}{2}}\\e^{\frac{i\xi_{+}}{2}}\end{array}\right)  +e^{-i\kappa x+i\theta_{-}}\left(\begin{array}{c}e^{\frac{-i\xi_{-}}{2}}\\ -e^{\frac{i\xi_{-}}{2}}\end{array}\right)
\right].
\end{align}
By substituting this ansatz into the energy functional and following straightforward calculation, one can obtain
\begin{align}
\mathcal{F}=&\frac{\rho_s}{8m}\left[2(\nabla\theta_{+})^2+2(\nabla \theta_{-})^2+\frac{1}{2}(\nabla\xi_{+})^2+\frac{1}{2}(\nabla\xi_{-})^2
+4\kappa(\partial_x\theta_{+}-\partial_x\theta_{-})
\right]
\nonumber \\ &
-\frac{\kappa\rho_s}{2m}\left[(\kappa+\partial_x\theta_{+})\cos\xi_{+}
+(\kappa-\partial_x\theta_{-})\cos\xi_{-}+\eta\partial_y\theta_{+}\sin\xi_{+}-\eta\partial_y\theta_{-}\sin\xi_{-}
\right]-\frac{g\rho_s^2}{8}\cos(\xi_{+}-\xi_{-}).
\end{align}
Here we have dropped some constants and the fast oscillating terms. Expanding this functional to the second order of $\theta$, $u$, $\xi_1\equiv\xi_{+}+\xi_{-}$ and $\xi_{2}\equiv\xi_{+}-\xi_{-}$, we obtain (up to some constants)
\begin{align}
\mathcal{F}=\frac{\rho_s}{8m}\left[4(\nabla\theta)^2+4(\nabla u)^2+\frac{1}{4}(\nabla\xi_1)^2+\frac{1}{4}(\nabla\xi_2)^2
\right]+\frac{\kappa\rho_s}{2m}\left[\frac{\kappa}{4}\xi_{1}^2+
\frac{\kappa}{4}\xi_{2}^2+\eta(\partial_y\theta)\xi_{2}+\eta(\partial_y u)\xi_{1}
\right]+\frac{g\rho_s^2}{16}\xi^2_2.
\end{align}
Noting that both $\xi_{1}$ and $\xi_{2}$ are massive modes, we can integrate them out and get the energy functional for $\theta$ and $u$ only
\begin{equation}
\mathcal{H}_{\text{eff}}=\frac{\rho_s}{2m}\left[(\nabla\theta)^2+(\nabla u)^2 - \eta^2 (\partial_y\theta) \frac{1}{1+\frac{mg\rho_s}{2\kappa^2}-\frac{\nabla^2}{4\kappa^2}}(\partial_y\theta) - \eta^2 (\partial_y u) \frac{1}{1-\frac{\nabla^2}{4\kappa^2}}(\partial_y u)\right]
\end{equation}
We can perform expansions in terms of $\nabla^2$ in the denominators. Then we can get
\begin{equation}
\mathcal{H}_{\text{eff}}= \mathcal{H}_{\text{eff}}^{\theta} + \mathcal{H}_{\text{eff}}^{u}
\end{equation}
where
\begin{equation}
\mathcal{H}^{\theta}_{\text{eff}}= \frac{\rho_s}{2m}\left[(\partial_x\theta)^2+ (1- \frac{\eta^2}{1+\frac{mg\rho_s}{2\kappa^2}})(\partial_y\theta)^2-\frac{\eta^2}{4\kappa^2(1+\frac{mg\rho_s}{2\kappa^2})^2}(\partial_y\theta)
\nabla^2(\partial_y\theta)+{\text{higher order terms.}}
\right]
\end{equation}
and
\begin{equation}
\mathcal{H}^{u}_{\text{eff}}= \frac{\rho_s}{2m}\left[(\partial_x u)^2+ (1-\eta^2)(\partial_y u)^2-\frac{\eta^2}{4\kappa^2}(\partial_y u)\nabla^2(\partial_y u)+{\text{higher order terms.}}
\right].
\end{equation}

It is clearly that all four terms $(\partial_x \theta)^2,~(\partial_y \theta)^2,~(\partial_x u)^2~\text{and} ~(\partial_y u)^2$ are preserved as long as $\eta<1$. With all these four terms present, other terms involving higher order derivatives of $\theta$ and $u$ (including $(\partial_y\theta)\nabla^2(\partial_y\theta)$ and $(\partial_y u)\nabla^2(\partial_y u)$) can be dropped in the long wave length limit we are considering.
Thus when $\eta<1$, we end up with an effective Hamiltonian for the stripe phase shown as follows:
\begin{align}
\mathcal{H}^{\text {ST}}_{\text{eff}}=\frac{\rho_s}{2m}\left\{(\partial_x\theta)^2+\left[(1-\eta^2)+\frac{\eta^2}{\alpha^2}\right](\partial_y\theta)^2 +(\partial_x u)^2+(1-\eta^2)(\partial_y u )^2\right\}, \label{effST}
\end{align}
where $\alpha=\sqrt{1+2\kappa^2/(m\rho_s g)}$ is defined in the paper.

In the isotropic limit, i.e. $\eta=1$, the terms $(\partial_x \theta)^2,~(\partial_y \theta)^2$ still exist, for which all other terms involving higher order derivatives of $\theta$ can be ignored. However, $(\partial_y u)^2$ vanishes exactly at the isotropic limit. Therefore, the term $(\partial_y u) \nabla^2(\partial_y u)$ must be kept. The $\mathcal{H}^{u}_{\text{eff}}$ part can be approximated as
\begin{equation}
\mathcal{H}^{u}_{\text{eff}}=\frac{\rho_s}{2m}\left[(\partial_x u)^2+\frac{1}{4\kappa^2}(\nabla\partial_y u)\cdot(\nabla\partial_y u)\right]=\frac{\rho_s}{2m}\left[(\partial_x u)^2+\frac{1}{4\kappa^2}(\partial_x\partial_y u)^2+\frac{1}{4\kappa^2}(\partial_y^2 u)^2\right],
\end{equation}
where an integral by part is performed. Here the second term $(\partial_x\partial_y u)^2$ can also been ignored in the long wave length limit because of the presence $(\partial_x u)^2$ term, and one can show that even we include the second term in our calculation, the behavior of the correlation functions of $u$ at large distance will not change qualitatively. Considering all of these, we arrive at the effective Hamiltonian in the isotropic limit.
\begin{align}
\mathcal{H}_{\text {eff}}^{{\text ST}}=\frac{\rho_s}{2m}\left[\left(\partial_x\theta\right)^2+\frac{\left(\partial_y\theta\right)^2}{\alpha^2}+
 \left(\partial_x u\right)^2+\frac{\left(\partial_y^2 u \right)^2}{4\kappa^2}
\right], \label{SWHeff}
\end{align}

For the plane wave phase, the ansatz introduced in the paper is
\begin{equation}
\varphi=\sqrt{\frac{\rho_s}{2}}e^{i\kappa x+i\theta}
\left(\begin{array}{c}
e^{-i\xi/2}\\
e^{i\xi/2}
\end{array}\right).\label{PWansatz}
\end{equation}
Similar to the stripe phase, by substituting the ansatz into the energy functional and following straightforward calculations, one can obtain
\begin{align}
\mathcal{F}=\frac{\rho_s}{2m}\left[(\nabla\theta)^2+\frac{(\nabla\xi)^2}{4}
-2\eta\kappa\xi(\partial_y\theta)+\kappa^2\xi^2\right]
\end{align}
Here we have dropped all the constants and fast oscillating terms, and expanded the functional to the second order of $\theta$ and $\xi$. Again, we find that $\xi$ is a massive field. By integrating out $\xi$, one gets a effective energy functional for $\theta$ only.
\begin{eqnarray}
\mathcal{H}_{\text {eff}} &=& \frac{\rho_s}{2m}\left[(\nabla\theta)^2-(\partial_y \theta)\frac{\eta^2}{1-\frac{\nabla^2}{4\kappa}}(\partial_y \theta)\right] \nonumber \\
& = &\frac{\rho_s}{2m}\left[(\partial_x\theta)^2+(1-\eta^2)(\partial_y\theta)^2-\frac{\eta^2}{4\kappa^2}(\partial _y \theta)\nabla^2(\partial _y \theta)+\text{higher order derivative terms.}
\right]\label{HeffPW}
\end{eqnarray}
By the similar argument presented for the stripe phase, only the first two terms in the bracket should be kept, as long as $\eta<1$. Therefore, our effective Hamiltonian for the plane phase in the anisotropic regime is
\begin{equation}
\mathcal{H}^{\text{PW}}_{\text{eff}}=\frac{\rho_s}{2m}\left[(\partial_x\theta)^2+(1-\eta^2)(\partial_y\theta)^2\right]
\end{equation}
In the isotropic limit, the $(\partial_y \theta)^2$ vanishes in Eq. \ref{HeffPW}. The third term in the second line of Eq. \ref{HeffPW} becomes important. Following the same reason given in discussing stripe phase, we obtain the effective Hamiltonian for the plane wave phase in the isotropic limit.
\begin{equation}
\mathcal{H}_{\text {eff}}^{^{\text PW}}=\frac{\rho_s}{2m}\left[\left(\partial_x\theta\right)^2+\frac{1}{4\kappa^2}(\partial_y^2\theta)^2\right]
.\label{PWHeff}
\end{equation}

\end{widetext}

\end{document}